\documentstyle[amssymb,floats,aps,prb,epsf]{revtex}
\epsfclipon 
\begin{document}

\draft
\title{Energy dependence of commensurate neutron scattering peak
in doped two-leg ladder antiferromagnet
Sr$_{14-x}$Ca$_{x}$Cu$_{24}$O$_{41}$}

\author{Jianhui He and Shiping Feng}
\address{Department of Physics and Key Laboratory of Beam
Technology and Material Modification, Beijing Normal University,
Beijing 100875, China\\
Interdisciplinary Center of Theoretical Studies, Chinese Academy
of Sciences, Beijing 100080, China\\
National Laboratory of Superconductivity, Chinese Academy of
Sciences, Beijing 100080, China}

\author{Wei Yeu Chen}
\address{Department of Physics, Tamkang University, Tamsui 25137,
Taiwan}

\maketitle

\begin{abstract}
The dynamical spin response of doped two-leg ladder
antiferromagnets is investigated based on the fermion-spin
approach. Our calculations clearly demonstrate a crossover from
the incommensurate antiferromagnetism in the weak interchain
coupling regime to commensurate spin fluctuation in the strong
interchain coupling regime. In particular, the nuclear spin-lattice
relaxation rate extracted from the commensurate spin fluctuation
decreases exponentially with decreasing temperatures. The behaviors
of the spin dynamics in the strong coupling regime are
quantitatively close to the experimental results of
Sr$_{14-x}$Ca$_{x}$Cu$_{24}$O$_{41}$.
\end{abstract}


In recent years the novel two-leg ladder antiferromagnet
Sr$_{14}$Cu$_{24}$O$_{41}$, being situated between doped one- and
two-dimensional antiferromagnets, have been experimentally
investigated as well as theoretically \cite{n1}. This followed
from the fact that this material has a spin liquid ground state
\cite{n101}, which may play a crucial role in the superconductivity
of doped cuprates as emphasized by Anderson \cite{n2}. Once
carriers are added to the material Sr$_{14}$Cu$_{24}$O$_{41}$, such
as the isovalent substitution of Ca for Sr, a metal-insulator
transition has been observed \cite{n1,n3,n5}, and further, this
doped two-leg ladder material Sr$_{14-x}$Ca$_{x}$Cu$_{24}$O$_{41}$
is a superconductor under pressure in low temperatures
\cite{n3,n5}. All cuprate superconductors found up now contain
square CuO$_{2}$ planes, whereas
Sr$_{14-x}$Ca$_{x}$Cu$_{24}$O$_{41}$ consists of two-leg ladders of
other Cu ions and edge-sharing CuO$_{2}$ chains \cite{n1,n3,n5},
and is the only known superconducting copper oxide without a square
lattice. Experimentally, it has been shown by virtue of transport
measurements that there are the regions of parameter space where
the resistivity is linear with temperatures \cite{n5}, one of the
hallmarks of the exotic normal state properties found in the
two-dimensional cuprates \cite{n4}. However, NMR and nuclear
quadrupole resonance (NQR), particularly inelastic neutron
scattering measurements in the same regions of parameter space
indicate that Sr$_{14-x}$Ca$_{x}$Cu$_{24}$O$_{41}$ in the normal
state is an antiferromagnet with the commensurate short-range
order \cite{n1,n6,n61}. This commensurate spin correlation is
energy dependent, and persists in a wide range of doping
\cite{n1,n6,n61}. Moreover, NMR and NQR spin-lattice relaxation
rate extracted from this commensurate spin fluctuation decreases
exponentially with decreasing temperatures \cite{n6,n61}. These
magnetic behaviors are different from these of the doped
two-dimensional high-T$_{c}$ cuprates \cite{n71,n8}, where the
incommensurate spin fluctuation for the single layer lanthanum
cuprate \cite{n71}, and both low-energy incommensurate spin
fluctuation and high-energy commensurate [$\pi$,$\pi$] resonance
for the bilayer yttrium cuprate \cite{n8} in the normal state are
observed.

Theoretically there is a general consensus that the unusual
physical properties of the two-leg ladder materials are due to the
quantum interference between the chains in the ladders \cite{n1}.
Applying the bosonization procedure to the two-leg $t$-$J$ and
Hubbard ladders, it is shown \cite{n111} that all spin excitations
are gapful and a singlet pairing becomes the dominant instability.
These results are confirmed by some research groups within the
different theoretical frameworks \cite{n112}. Furthermore, based
on the antiferromagnetic Heisenberg model, the susceptibility and
spin-lattice relaxation rate have been discussed \cite{n91}, and
the results show that the large contribution to the spin-lattice
relaxation rate comes from processes with the wave vectors around
the antiferromagnetic zone center. Within the one dimensional
gapped quantum nonlinear $\sigma$ model, an effective classical
model has been developed to study the spin transport of the two-leg
ladder antiferromagnets \cite{n92}, and the result obtained for the
spin-lattice relaxation rate is close to what is experimentally
measured. However, to the best of our knowledge, no systematic
calculations have been performed to show why the commensurate
neutron scattering peak can be observed in the doped two-leg ladder
antiferromagnet Sr$_{14-x}$Ca$_{x}$Cu$_{24}$O$_{41}$? This is a
challenge issue since it is closely related to the doped Mott
insulating state that forms the basis for the superconductivity
\cite{n3}. In this paper, using the fermion-spin theory \cite{n10}
which implements properly the local single occupancy constraint, we
calculate explicitly the dynamical spin structure factor within a
$t$-$J$ ladder and show that in the regions of parameter space
given by the experiments, one can reproduce all main magnetic
features observed experimentally on
Sr$_{14-x}$Ca$_{x}$Cu$_{24}$O$_{41}$ in the normal state, including
the energy dependence of the neutron scattering peak position and
exponential decrease of the nuclear spin-lattice relaxation rate.

The two-leg ladder is defined as two parallel chains of ions, with
bonds among them such that the interchain coupling is comparable
in strength to the couplings along the chains, while the coupling
between the two chains that participates in this structure is
through rungs \cite{n1,n3}. The essential properties of the doped
two-leg ladder antiferromagnet can be described by the $t$-$J$
ladder as \cite{n95},
\begin{eqnarray}
H&=&-t_{\parallel}\sum_{i\hat{\eta}a\sigma}C_{ia\sigma}^{\dagger}
C_{i+\hat{\eta}a\sigma}-t_{\perp}\sum_{i\sigma}
(C_{i1\sigma}^{\dagger}C_{i2\sigma}+{\rm h.c.}) \nonumber \\
&-&\mu\sum_{ia\sigma}C_{ia\sigma }^{\dagger}C_{ia\sigma }+
J_{\parallel}\sum_{i\hat{\eta}a}{\bf S}_{ia}\cdot
{\bf S}_{i+\hat{\eta}a}+J_{\perp}\sum_{i}{\bf S}_{i1}\cdot
{\bf S}_{i2},
\end{eqnarray}
with the local constraint
$\sum_{\sigma}C_{ia\sigma}^{\dagger}C_{ia\sigma}\leq 1$ to remove
double occupancy of any site, where $\hat{\eta}=\pm c_{0}\hat{x}$,
$c_{0}$ is the lattice constant of the two-leg ladder lattice,
which is set as the unit hereafter, $i$ runs over all rungs,
$\sigma(=\uparrow,\downarrow)$ and $a(=1,2)$ are spin and leg
indices, respectively, $C^{\dagger}_{ia\sigma}$ ($C_{ia\sigma}$)
are the electron creation (annihilation) operators,
${\bf S}_{ia}=C^{\dagger}_{ia}\vec{\sigma}C_{ia}/2$ are the spin
operators with $\vec{\sigma}=(\sigma_{x},\sigma_{y},\sigma_{z})$
as the Pauli matrices, and $\mu$ is the chemical potential. For
the two-leg ladder materials, it has been shown from experiments
\cite{n1,n6,n61} that the exchange coupling $J_{\parallel}$
along the legs is greater than exchange coupling $J_{\perp}$
across a rung, {\it i.e.}, $J_{\parallel}>J_{\perp}$, and
similarly the hopping $t_{\parallel}$ along the legs is greater
than the rung hopping strength $t_{\perp}$, {\it i.e.},
$t_{\parallel}>t_{\perp}$. On the other hand, the strong electron
correlation in the $t$-$J$ model is reflected by the local
constraint \cite{n2}, which can be treated properly within the
fermion-spin theory \cite{n10}, where electron operators
$C_{ia\uparrow}=h^{\dagger}_{ia}S^{-}_{ia}$ and
$C_{ia\downarrow}=h^{\dagger}_{ia}S^{+}_{ia}$ are represented by
the spinless fermion operator $h_{ia}$ carrying the charge (holon)
and the pseudospin operator $S_{ia}$ representing the spin
(spinon), then it naturally incorporates the physics of charge-spin
separation. In this case, the low-energy behavior of the $t$-$J$
ladder (1) can be rewritten in the fermion-spin representation as,
\begin{eqnarray}
H&=&t_{\parallel}\sum_{i\hat{\eta}a}h^{\dagger}_{i+\hat{\eta}a}
h_{ia}(S^{+}_{ia}S^{-}_{i+\hat{\eta}a}+S^{-}_{ia}
S^{+}_{i+\hat{\eta}a})+t_{\perp}\sum_{i}(h^{\dagger}_{i1}h_{i2}+
h^{\dagger}_{i2}h_{i1})(S^{+}_{i1}S^{-}_{i2}+S^{-}_{i1}S^{+}_{i2})
\nonumber \\
&+&\mu\sum_{ia}h^{\dagger}_{ia}h_{ia}+J_{\parallel {\rm eff}}
\sum_{i\hat{\eta}a}{\bf S}_{ia}\cdot {\bf S}_{i+\hat{\eta}a}
+J_{\perp{\rm eff}}\sum_{i}{\bf S}_{i1}\cdot {\bf S}_{i2},
\end{eqnarray}
where $J_{\parallel {\rm eff}}=J_{\parallel}[(1-p)^{2}-
\phi_{\parallel}^{2}]$, $J_{\perp{\rm eff}}=J[(1-p)^{2}-
\phi^{2}_{\perp}]$, $p$ is the hole doping concentration, the
holon particle-hole order parameters $\phi_{\parallel}=\langle
h^{\dagger}_{ia}h_{i+\hat{\eta}a}\rangle$, $\phi_{\perp}=\langle
h^{\dagger}_{i1}h_{i2}\rangle$, and $S^{+}_{ia}$ ($S^{-}_{ia}$)
as the pseudospin raising (lowering) operators. It has been shown
\cite{n10} that the constrained electron operator can be mapped
exactly using the fermion-spin transformation defined with an
additional projection operator. However, this projection operator
is cumbersome to handle in the actual calculations, and we have
dropped it in Eq. (2) and in the subsequent calculation. It has
also been shown \cite{n10} that such treatment leads to errors of
the order $p$ in counting the number of spin states, which is
negligible for small dopings. In the two-leg ladder systems,
because of the two coupled chains, the energy spectrum has two
branches, therefore the one-particle spinon and holon Green's
functions are matrices, and can be expressed as
$D(i-j,\tau-\tau^{\prime})=D_{L}(i-j,\tau-\tau^{\prime})+\sigma_{x}
D_{T}(i-j,\tau-\tau^{\prime})$ and $g(i-j,\tau-\tau^{\prime})=
g_{L}(i-j,\tau-\tau^{\prime})+\sigma_{x}g_{T}
(i-j,\tau-\tau^{\prime})$, respectively, where the longitudinal
and transverse parts are defined as $D_{L}(i-j,\tau-\tau^{\prime})=
-\langle T_{\tau}S_{ia}^{+}(\tau)S_{ja}^{-}(\tau^{\prime})\rangle$,
$g_{L}(i-j,\tau-\tau^{\prime})=-\langle T_{\tau}h_{ia}(\tau)
h_{ja}^{\dagger}(\tau^{\prime})\rangle$ and $D_{T}
(i-j,\tau-\tau^{\prime})=-\langle T_{\tau}S_{ia}^{+}(\tau)
S_{ja^{\prime}}^{-}(\tau^{\prime})\rangle$,
$g_{T}(i-j,\tau-\tau^{\prime})=-\langle T_{\tau}h_{ia}(\tau)
h_{ja^{\prime}}^{\dagger}(\tau^{\prime})\rangle$ with $a\neq
a^{\prime}$. Within the charge-spin separation, the spin
fluctuation couples only to spinons \cite{n99}, however, the
strong correlation between holons and spinons still is included
self-consistently through the holon's order parameters entering the
spinon's propagator, then both holons and spinons are responsible
for the spin dynamics. In this case, the spin dynamics of the doped
square lattice antiferromagnet have been discussed \cite{n11}, and
the results are consistent with the experiments \cite{n71,n8}.
Following their discussions \cite{n11}, the dynamical spin
structure factor of the doped two-leg ladder antiferromagnet is
obtained explicitly as,
\begin{eqnarray}
S({\bf k},\omega)&=&-2[1+n_{B}(\omega)][2{\rm Im}D_{L}({\bf k},
\omega)+2{\rm Im}D_{T}({\bf k},\omega)] \nonumber \\
&=&-{4[1+n_{B}(\omega)](B^{(1)}_{k})^{2}{\rm Im}
\Sigma_{LT}^{(s)}({\bf k},\omega)\over[\omega^{2}-
(\omega^{(1)}_{k})^{2}-B^{(1)}_{k}{\rm Re}\Sigma_{LT}^{(s)}
({\bf k},\omega)]^{2}+[B^{(1)}_{k}{\rm Im}\Sigma_{LT}^{(s)}
({\bf k},\omega)]^{2}},
\end{eqnarray}
where the full spinon Green's function, $D^{-1}({\bf k},\omega)=
D^{(0)-1}({\bf k},\omega)-\Sigma^{(s)}({\bf k},\omega)$, with
the mean-field spinon Green's function,
\begin{eqnarray}
D^{(0)}_{L}({\bf k},\omega)&=&{1\over 2}\sum_{\nu}{B^{(\nu)}_{k}
\over \omega^{2}-(\omega^{(\nu)}_{k})^{2}}, \nonumber \\
D^{(0)}_{T}({\bf k},\omega)&=&{1\over 2}\sum_{\nu}(-1)^{\nu+1}
{B^{(\nu)}_{k}\over \omega^{2}-(\omega^{(\nu)}_{k})^{2}},
\end{eqnarray}
$\nu=1, 2$, ${\rm Im}\Sigma_{LT}^{(s)}({\bf k},\omega)={\rm Im}
\Sigma_{L}^{(s)}({\bf k},\omega)+{\rm Im}\Sigma_{T}^{(s)}({\bf k},
\omega)$, ${\rm Re}\Sigma_{LT}^{(s)}({\bf k},\omega)={\rm Re}
\Sigma_{L}^{(s)}({\bf k},\omega)+{\rm Re}\Sigma_{T}^{(s)}({\bf k},
\omega)$, while ${\rm Im}\Sigma^{(s)}_{L}({\bf k},\omega)$
(${\rm Im}\Sigma^{(s)}_{T}({\bf k},\omega)$) and ${\rm Re}
\Sigma^{(s)}_{L}({\bf k},\omega)$ (${\rm Re}\Sigma^{(s)}_{T}
({\bf k},\omega)$) are the imaginary and real parts of the second
order longitudinal (transverse) spinon self-energy, respectively,
obtained from the holon bubble as $\Sigma_{L}^{(s)}({\bf k},\omega)
=(1/L)^{2}\sum_{pp'}\sum_{\nu\nu'\nu''}\Pi_{\nu\nu'\nu''}({\bf k},
{\bf p},{\bf p'},\omega)$ and $\Sigma_{T}^{(s)}({\bf k},\omega)=
(1/L)^{2}\sum_{pp'}\sum_{\nu\nu'\nu''}(-1)^{\nu+\nu'+\nu''+1}
\Pi_{\nu\nu'\nu''}({\bf k},{\bf p},{\bf p'},\omega)$ with $L$ is
the number of rungs, and
\begin{eqnarray}
\Pi_{\nu\nu'\nu''}({\bf k},{\bf p},{\bf p'},\omega)&=&\left
(2t_{\parallel}[{\rm cos}(p'+p+k)+{\rm cos}(p'-k)]+t_{\perp}
[(-1)^{\nu'+\nu''}+(-1)^{\nu+\nu''}]\right )^{2} \nonumber \\
&\times&{B^{(\nu'')}_{k+p}\over 16\omega^{(\nu'')}_{k+p}}
\left ({F^{(1)}_{\nu\nu'\nu''}({\bf k},{\bf p},{\bf p'})
\over\omega+\xi^{(\nu')}_{p+p'}-\xi^{(\nu)}_{p'}-
\omega^{(\nu'')}_{k+p}}-{F^{(2)}_{\nu\nu'\nu''}
({\bf k},{\bf p},{\bf p'})\over \omega+\xi^{(\nu')}_{p+p'}
-\xi^{\nu}_{p'}+\omega^{(\nu'')}_{k+p}}\right ),
\end{eqnarray}
where $B^{(\nu)}_{k}=B_{k}-J_{\perp {\rm eff}}[\chi_{\perp}+2
\chi^{z}_{\perp}(-1)^{\nu}][\epsilon_{\perp}+(-1)^{\nu}]$,
$B_{k}=\lambda[(2\epsilon_{\parallel}\chi^{z}_{\parallel}+
\chi_{\parallel}){\rm cos}k-(\epsilon_{\parallel}\chi_{\parallel}
+2\chi^{z}_{\parallel})]$, $\lambda=4J_{{\parallel}{\rm eff}}$,
$\epsilon_{\parallel}=1+2t_{\parallel}\phi_{\parallel}/
J_{{\parallel}{\rm eff}}$, $\epsilon_{\perp}=1+4t_{\perp}
\phi_{\perp}/J_{\perp{\rm eff}}$, and
\begin{eqnarray}
F^{(1)}_{\nu\nu'\nu''}({\bf k},{\bf p},{\bf p'}) &=& n_{F}
(\xi^{(\nu')}_{p+p'})[1-n_{F}(\xi^{(\nu)}_{p'})]-n_{B}
(\omega^{(\nu'')}_{k+p})[n_{F}(\xi^{(\nu)}_{p'})-n_{F}
(\xi^{(\nu')}_{p+p'})], \nonumber \\
F^{(2)}_{\nu\nu'\nu''}({\bf k},{\bf p},{\bf p'}) &=& n_{F}
(\xi^{(\nu')}_{p+p'})[1-n_{F}(\xi^{(\nu)}_{p'})]+[1+n_{B}
(\omega^{(\nu'')}_{k+p})][n_{F}(\xi^{(\nu)}_{p'})-n_{F}
(\xi^{(\nu')}_{p+p'})],
\end{eqnarray}
$n_{F}(\xi^{(\nu)}_{k})$ and $n_{B}(\omega^{(\nu)}_{k})$ are the
fermion and boson distribution functions, respectively, the
mean-field holon excitations $\xi^{(\nu)}_{k}=4t_{\parallel}
\chi_{\parallel}{\rm cos}k+\mu+2\chi_{\perp}t_{\perp}(-1)^{\nu+1}$,
and the mean-field spinon excitations, $(\omega^{(\nu)}_{k})^{2}=
\omega^{2}_{k}+\Delta^{2}_{k}(-1)^{\nu+1}$ with $\omega^{2}_{k}=
A_{1}({\rm cos}k)^{2}+A_{2}{\rm cos}k+A_{3}$, $\Delta^{2}_{k}=
X_{1}{\rm cos}k+X_{2}$, where
\begin{eqnarray}
A_{1}&=&\alpha\epsilon_{\parallel}\lambda^{2}(\chi_{\parallel}/2+
\epsilon_{\parallel}\chi^{z}_{\parallel}), \nonumber \\
A_{2}&=&-\epsilon_{\parallel}\lambda^{2}[\alpha
(\epsilon_{\parallel}\chi_{\parallel}/2+\chi^{z}_{\parallel})/2-
\alpha(C^{z}_{\parallel}+C_{\parallel}/2)-(1-\alpha)/4]
\nonumber \\
&-&\alpha\lambda J_{\perp{\rm eff}}[\epsilon_{\parallel}
(C^{z}_{\perp}+\chi^{z}_{\perp})+\epsilon_{\perp}(C_{\perp}+
\epsilon\chi_{\perp})/2], \nonumber \\
A_{3}&=&\lambda^{2}[\alpha(C^{z}_{\parallel}+
\epsilon^{2}_{\parallel}C_{\parallel}/2)+(1-\alpha)
(1+\epsilon^{2}_{\parallel})/8-\alpha\epsilon_{\parallel}
(\chi_{\parallel}/2+\epsilon_{\parallel}\chi^{z}_{\parallel})/2]
\nonumber \\
&+&\alpha\lambda J_{\perp{\rm eff}}[\epsilon_{\parallel}
\epsilon_{\perp}C_{\perp}+2C^{z}_{\perp}]+J^{2}_{\perp{\rm eff}}
(\epsilon^{2}_{\perp}+1)/4, \nonumber \\
X_{1}&=&\alpha\lambda J_{\perp{\rm eff}}[(\epsilon_{\perp}
\chi_{\parallel}+\epsilon_{\parallel}\chi_{\perp})/2+
\epsilon_{\parallel}\epsilon_{\perp}(\chi^{z}_{\perp}+
\chi^{z}_{\parallel})], \nonumber \\
X_{2}&=&-\alpha\lambda J_{\perp{\rm eff}}[\epsilon_{\parallel}
\epsilon_{\perp}\chi_{\parallel}/2+\epsilon_{\perp}
(\chi^{z}_{\parallel}+C^{z}_{\perp})+\epsilon_{\parallel}
C_{\perp}/2]-\epsilon_{\perp}J^{2}_{\perp{\rm eff}}/2,
\end{eqnarray}
with the spinon correlation functions $\chi_{\parallel}=\langle
S_{ai}^{+}S_{ai+\hat{\eta}}^{-}\rangle$, $\chi^{z}_{\parallel}=
\langle S_{ai}^{z}S_{i+\hat{\eta}}^{z}\rangle$, $\chi_{\perp}=
\langle S_{1i}^{+}S_{2i}^{-}\rangle$, $\chi^{z}_{\perp}=\langle
S_{1i}^{z}S_{2i}^{z}\rangle$, $C_{\parallel}=(1/4)
\sum_{\hat{\eta}\hat{\eta'}}\langle S_{ai+\hat{\eta}}^{+}
S_{ai+\hat{\eta'}}^{-}\rangle$, $C^{z}_{\parallel}=(1/4)
\sum_{\hat{\eta}\hat{\eta'}}\langle S_{ai+\hat{\eta}}^{z}
S_{ai+\hat{\eta'}}^{z}\rangle$, $C_{\perp}=(1/2)\sum_{\hat{\eta}}
\langle S_{2i}^{+}S_{1i+\hat{\eta}}^{-}\rangle$, and $C^{z}_{\perp}
=(1/2)\sum_{\hat{\eta}}\langle S_{1i}^{z}S_{2i+\hat{\eta}}^{z}
\rangle$. In order to satisfy the sum rule for the correlation
function $\langle S^{+}_{ai}S^{-}_{ai}\rangle =1/2$ in the absence
of the antiferromagnetic long-range-order, a decoupling parameter
$\alpha$ has been introduced in the mean-field calculation, which
can be regarded as the vertex correction \cite{n96}. All the
mean-field order parameters, decoupling parameter, and chemical
potential have been determined self-consistently, as done in the
two-dimensional case \cite{n11,n96}.

\begin{figure}[prb]
\epsfxsize=2.5in\centerline{\epsffile{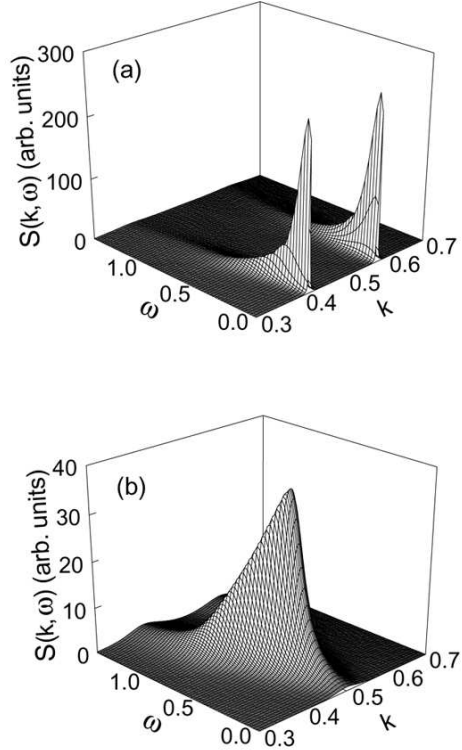}}
\caption{The dynamical spin structure factor in the ($k,\omega$)
plane at $p=0.16$ with $T=0.05J_{\parallel}$ for
$t_{\parallel}/J_{\parallel}=2.5$, (a) $t_{\perp}/t_{\parallel}=
0.5$ and $J_{\perp}/J_{\parallel}=0.5$,
and (b) $t_{\perp}/t_{\parallel}=0.70$ and
$J_{\perp}/J_{\parallel}=0.70$.}
\end{figure}
\begin{figure}[prb]
\epsfxsize=2.5in\centerline{\epsffile{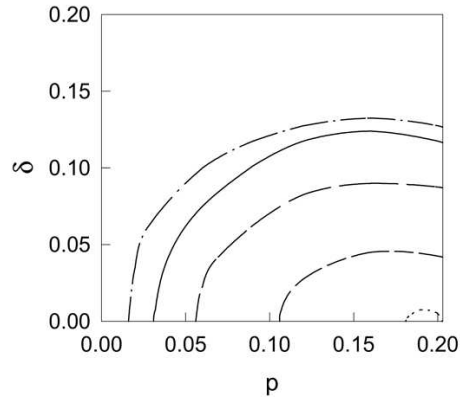}}
\caption{The doping dependence of incommensurability $\delta(x)$
in $T=0.05J_{\parallel}$ and $\omega=0.4J_{\parallel}$ for
$t_{\parallel}/J_{\parallel}=2.5$, $t_{\perp}/t_{\parallel}=0$ and
$J_{\perp}/J_{\parallel}=0$ (dash-dotted line), $t_{\perp}/
t_{\parallel}=0.3$ and $J_{\perp}/J_{\parallel}=0.3$ (solid line),
$t_{\perp}/t_{\parallel}=0.5$ and $J_{\perp}/J_{\parallel}=0.5$
(long-dashed line), $t_{\perp}/t_{\parallel}=0.65$ and
$J_{\perp}/J_{\parallel}=0.65$ (short-dashed line), and
$t_{\perp}/t_{\parallel}=0.69$ and $J_{\perp}/J_{\parallel}=0.69$
(dotted line). }
\end{figure}

In Fig. 1, we represent the dynamical spin structure factor
$S({\bf k},\omega)$ in the ($k,\omega$) plane at doping $p=0.16$
with temperature $T=0.05J_{\parallel}$ for
$t_{\parallel}/J_{\parallel}=2.5$, (a)
$t_{\perp}/t_{\parallel}=0.5$, $J_{\perp}/J_{\parallel}=0.5$, and
(b) $t_{\perp}/t_{\parallel}=0.7$, $J_{\perp}/J_{\parallel}=0.7$,
hereafter we use the units of $[2\pi]$. Obviously, an interchain
coupling dependence of the incommensurate-commensurate transition
occurs. To check this point explicitly, the calculated dynamical
spin structure factor spectrum has been used to extract the doping
and interchain coupling dependence of the incommensurability
$\delta$, defined as the deviation of the peak position from the
antiferromagnetic wave vector ${\bf Q}=[1/2]$, and the results in
$T=0.05J_{\parallel}$ and $\omega=0.4 J_{\parallel}$ for
$t_{\parallel}/J_{\parallel}=2.5$, $t_{\perp}/t_{\parallel}=0$ and
$J_{\perp}/J_{\parallel}=0$ (dash-dotted line),
$t_{\perp}/t_{\parallel}=0.3$ and $J_{\perp}/J_{\parallel}=0.3$
(solid line), $t_{\perp}/t_{\parallel}=0.5$ and
$J_{\perp}/J_{\parallel}=0.5$ (long-dashed line),
$t_{\perp}/t_{\parallel}=0.65$ and $J_{\perp}/J_{\parallel}=0.65$
(short-dashed line), and $t_{\perp}/t_{\parallel}=0.69$ and
$J_{\perp}/J_{\parallel}=0.69$ (dotted line) are shown in Fig. 2.
We therefore find that there are a {\it critical} values of
$t_{\perp}/t_{\parallel}=0.7$ and $J_{\perp}/J_{\parallel}=0.7$ in
$T=0.05J_{\parallel}$ and $\omega=0.4J_{\parallel}$. In the weak
coupling regime with $t_{\perp}/t_{\parallel}\ll 0.7$ and
$J_{\perp}/J_{\parallel}\ll 0.7$, the commensurate scattering peak
near the half-filling is split into two peaks at
$[(1\pm\delta)/2]$, where $\delta$ increases initially with the
hole concentration in lower dopings, but it saturates at higher
dopings. In this case, spinons and holons are more likely to move
along the legs of the ladders, rendering the materials
quasi-one-dimension. However, the range of the incommensurate spin
correlation decreases with increasing the strength of the
interchain coupling, and the commensurate spin fluctuation appears
in the whole doping range in the strong coupling regime with
$t_{\perp}/t_{\parallel}\geq 0.7$ and
$J_{\perp}/J_{\parallel}\geq 0.7$. Many experimental analyses
\cite{n1,n5,n6} have indicated that
$J_{\perp}/J_{\parallel} > 0.7$ for the doped two-leg ladder
antiferromagnet Sr$_{14-x}$Ca$_{x}$Cu$_{24}$O$_{41}$, this is why
that the incommensurate spin correlation is not observed in
Sr$_{14-x}$Ca$_{x}$Cu$_{24}$O$_{41}$ from the inelastic neutron
scattering experiments \cite{n1,n5,n6}. For a better understanding
of the evolution of the commensurate scattering peak with energy
in Sr$_{14-x}$Ca$_{x}$Cu$_{24}$O$_{41}$, we have made a series of
scans for the dynamical spin structure factor $S({\bf k},\omega)$
at different dopings, and the result for $p=0.2$ with
$T=0.05J_{\parallel}$ for $t_{\parallel}/J_{\parallel}=2.5$,
$t_{\perp}/t_{\parallel}=0.77$ and $J_{\perp}/J_{\parallel}=0.77$
is plotted in Fig. 3 in comparison with the experimental data
\cite{n6} taken on Sr$_{14-x}$Ca$_{x}$Cu$_{24}$O$_{41}$ (inset)
with $x=11.5$ ($p\approx 0.2$) and $J_{\perp}/J_{\parallel}=0.77$.
This result shows that the commensurate spin fluctuation is energy
dependent, and the commensurate scattering peak, which is similar
to the resonance peak in the bilayer cuprate in the normal state
\cite{n8}, is located at energy $\omega=0.93 J_{\parallel}$. This
reflects that the anticipated spin gap $\Delta_{S}=0.93
J_{\parallel}\approx 83.7$meV (Ref. \cite{n6}) is larger than the
spin gap $\approx 32.1$meV observed \cite{n6} in
Sr$_{14-x}$Ca$_{x}$Cu$_{24}$O$_{41}$, which may mean that the
simplest two-leg $t$-$J$ ladder can not be regarded as the
complete model for the quantitative comparison with the
doped two-leg ladder antiferromagnet. Furthermore, we have also
made a series of scans for the dynamical spin structure factor at
different temperatures, and found that the weight of the peak is
suppressed severely with increasing temperatures. Our present
results are in qualitative agreement with the experimental data
\cite{n6}.  On the other hand, our present conclusion in the strong
coupling regime contradict the numerical result of the isotropic
two-leg Hubbard ladder \cite{n19}, where the equal-time spin
structure factor has been calculated using the density-matrix
renormalization group method, and the result shows that the
commensurate peak at the half-filling is split into two peaks with
dopings. But we want to stress that these results are obtained
within the two-leg Hubbard ladder, where the Hubbard $U$ is finite.
However, in the $t$-$J$ model, the doubly occupied Hilbert space
has been pushed to infinity as the Hubbard $U\rightarrow \infty$,
and therefore the dynamical spin structure factor in the $t$-$J$
model only describes the lower Hubbard band. This may lead to some
different results between the Hubbard and $t$-$J$ models. Of cause,
this has to be checked by the further density-matrix
renormalization group study for the $t$-$J$ ladder.

\begin{figure}[prb]
\epsfxsize=2.5in\centerline{\epsffile{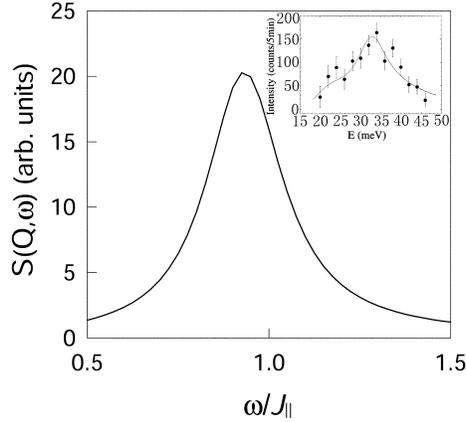}}
\caption{The dynamical spin structure factor at the AF wave vector
${\bf Q}=[1/2]$ for $p=0.2$ with $T=0.05J_{\parallel}$ for
$t_{\parallel}/J_{\parallel}=2.5$, $t_{\perp}/t_{\parallel}=0.77$
and $J_{\perp}/J_{\parallel}=0.77$. Inset: the experimental result
\cite{n6} on Sr$_{14-x}$Ca$_{x}$Cu$_{24}$O$_{41}$ with $x=11.5$
($p\approx 0.2$) and $J_{\perp}/J_{\parallel}=0.77$ taken from
Ref. \protect\cite{n6}.}
\end{figure}
\begin{figure}[prb]
\epsfxsize=2.5in\centerline{\epsffile{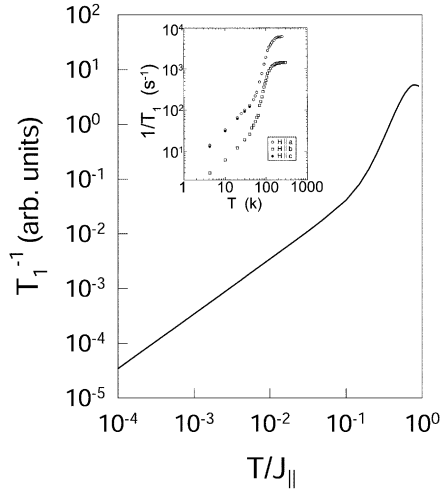}}
\caption{The temperature dependence of spin-lattice relaxation
time $1/T_{1}$ in both logarithmic scales at $p=0.20$ for
$t_{\parallel}/J_{\parallel}=2.5$, $t_{\perp}/t_{\parallel}=0.77$
and $J_{\perp}/J_{\parallel}=0.77$. Inset: the experimental result
on Sr$_{14-x}$Ca$_{x}$Cu$_{24}$O$_{41}$ with $x=11.5$
($p\approx 0.2$) and $J_{\perp}/J_{\parallel}=0.77$ taken from
Ref. \protect\cite{n61}.}
\end{figure}

One of the most important features of the spin dynamics in the
doped two-leg ladder antiferromagnet is the spin-lattice relaxation
time $1/T_{1}$ \cite{n61}. This spin-lattice relaxation time is
closely related with the dynamical spin structure factor, and can
be expressed as,
\begin{eqnarray}
1/T_{1}=2K_{B}T/(g^{2}\mu^{2}_{B}\hbar)\lim
\limits_{\omega\rightarrow 0}(1/N)\sum_{k}F^{2}_{\alpha}({\bf k})
\chi^{\prime\prime}({\bf k},\omega)/\omega,
\end{eqnarray}
with $g$ is the $g$ factor, $\mu_{0}$ is the Bohr magneton,
$F_{\alpha}({\bf k})$ is the form factors, and dynamical spin
susceptibility $\chi^{\prime\prime}({\bf k},\omega)$ is related to
the dynamical spin structure factor by the fluctuation-dissipation
theorem as, $\chi^{\prime\prime}({\bf k},\omega)=
(1-e^{-\beta\omega})S({\bf k},\omega)$. The form factors have
dimension of energy, and magnitude determined by atomic physics,
and $k$ dependence determined by geometry. Since the strong
short-range commensurate spin fluctuation in the strong coupling
regime in the dynamical spin structure factor (3), the main
contribution to $1/T_{1}$ comes from the region around the
antiferromagnetic wave vector, therefore we can set
$F_{\alpha}({\bf k})$ as constant without loss the generality
\cite{n91}. In this case, the spin-lattice relaxation time
$1/T_{1}$ has been evaluated and the result at $p=0.20$ for
$t_{\parallel}/J_{\parallel}=2.5$, $t_{\perp}/t_{\parallel}=0.77$
and $J_{\perp}/J_{\parallel}=0.77$ is plotted in Fig. 4 in
comparison with the experimental data \cite{n61} taken on
Sr$_{14-x}$Ca$_{x}$Cu$_{24}$O$_{41}$ (inset) with $x=11.5$
($p\approx 0.2$) and $J_{\perp}/J_{\parallel}=0.77$, where we have
chosen units $\hbar=K_{B}=1$. This result indicates that in low
temperatures the spin-lattice relaxation time decreases
exponentially with decreasing temperatures, in agreement with the
experiments \cite{n61}.

The dynamical spin structure factor in Eq. (3) has a well-defined
resonance character, where $S({\bf k},\omega)$ exhibits peaks when
the incoming neutron energy $\omega$ is equal to the renormalized
spin excitation $E^{2}_{k}=(\omega^{(1)}_{k})^{2}+ B^{(1)}_{k}
{\rm Re}\Sigma^{(s)}_{LT}({\bf k},E_{k})$, {\it i.e.},
$W({\bf k}_{c},\omega)\equiv [\omega^{2}-(\omega^{(1)}_{k_{c}})^{2}
-B^{(1)}_{k_{c}}{\rm Re}\Sigma^{(s)}_{LT}({\bf k}_{c},\omega)]^{2}
=(\omega^{2}-E^{2}_{k_{c}})^{2}\sim 0$ for certain critical wave
vectors ${\bf k}_{c}$. Then the height of these peaks is determined
by the imaginary part of the spinon self-energy $1/{\rm Im}
\Sigma^{(s)}_{LT}({\bf k}_{c},\omega)$. This renormalized spin
excitation is doping, energy, and interchain coupling dependent.
In the present spinon self-energy ${\rm Re}\Sigma_{LT}^{(s)}
({\bf k},\omega)={\rm Re}\Sigma_{L}^{(s)}({\bf k},\omega)+{\rm Re}
\Sigma_{T}^{(s)}({\bf k},\omega)$, ${\rm Re}\Sigma_{L}^{(s)}
({\bf k},\omega)<0$ favors the one-dimensional behaviors, while
${\rm Re}\Sigma_{T}^{(s)}({\bf k},\omega)>0$ characterizes the
quantum interference between the chains in the ladders, therefore
there is a competition between ${\rm Re}\Sigma_{L}^{(s)}({\bf k},
\omega)$ and ${\rm Re}\Sigma_{T}^{(s)}({\bf k},\omega)$. In the
weak coupling regime, the main contribution for ${\rm Re}
\Sigma_{LT}^{(s)}({\bf k},\omega)$ may come from ${\rm Re}
\Sigma_{L}^{(s)}({\bf k},\omega)$, and spinons and holons are more
likely to move along the legs, then the incommensurate spin
correlation emerges, where the essential physics is almost the same
as in the two-dimensional $t$-$J$ model \cite{n11}. Near the
half-filling, the spin excitations are centered around the
antiferromagnetic wave vector $[1/2]$, so the commensurate
antiferromagnetic peak appears there. Upon doping, the holes
disturb the antiferromagnetic background. Within the fermion-spin
framework, as a result of self-consistent motion of holons and
spinons, the incommensurate spin correlation is developed beyond
certain critical doping, which means, the low-energy spin
excitations drift away from the antiferromagnetic wave vector,
where the physics is dominated by the spinon self-energy
${\rm Re}\Sigma_{L}^{(s)}({\bf k},\omega)$ renormalization due to
holons. However, the quantum interference effect between the chains
manifests itself by the interchain coupling, {\it i.e.}, this
quantum interference increases with increasing the strength of the
interchain coupling. Thus in the strong coupling regime, ${\rm Re}
\Sigma_{T}^{(s)}({\bf k},\omega)$ may cancel the most
incommensurate spin correlation contributions from ${\rm Re}
\Sigma_{L}^{(s)}({\bf k},\omega)$, then the commensurate spin
fluctuation appears. In this sense, the interchain coupling is a
crucial role to determine the symmetry of the spin fluctuation in
the doped two-leg ladder antiferromagnet. Since the height of the
peaks is determined by damping, it is fully understandable that
the weight of the peak is suppressed as the temperature are
increased.

To conclude we have discussed the spin dynamics of the doped
two-leg ladder antiferromagnet within the $t$-$J$ ladder. Our
calculations clearly show a crossover from the incommensurate spin
correlation to commensurate spin fluctuation characterized by the
rung to chain coupling. However, in the regions of parameter space
given by experiments, the $t$-$J$ ladder can correctly reproduce
all main magnetic features of the doped two-leg ladder
antiferromagnet Sr$_{14-x}$Ca$_{x}$Cu$_{24}$O$_{41}$, including
the energy dependence of the neutron scattering peak position and
exponential decrease of the nuclear spin-lattice relaxation rate.

\acknowledgments
The authors would like to thank Dr. F. Yuan and Professor X. Wang
for helpful discussions. This work was supported by the National
Natural Science Foundation under the Grant Nos. 10074007, 10125415,
and 90103024, the special funds from the Ministry of Science and
Technology of China, and the National Science Council under Grant
No. NSC 90-2816-M-032-0001-6.

\end{document}